\documentclass[prc,twocolumn,showpacs,showkeys,nofootinbib,superscriptaddress]{revtex4}
\usepackage{graphicx}
\usepackage{dcolumn}
\usepackage{epsfig}
\usepackage{bm}
\usepackage{xcolor}

\begin{document}

\title{Muon capture rates: Evaluation within the Quasiparticle Random Phase Approximation}

\author{Fedor~\v{S}imkovic}
\email{simkovic@teller.dnp.fmph.uniba.sk}
\affiliation{Faculty of Mathematics, Physics and Informatics, Comenius University in Bratislava, 842~48~Bratislava, Slovakia}
\affiliation{Bogoliubov Laboratory of Theoretical Physics, Joint Institute for Nuclear Research, 141980~Dubna, Russia}
\affiliation{Institute of Experimental and Applied Physics, Czech Technical University in Prague, 110~00~Prague, Czech Republic}
\author{Rastislav Dvornick\'{y}}
\email{rastonator@gmail.com}
\affiliation{Faculty of Mathematics, Physics and Informatics, Comenius University in Bratislava, 842~48~Bratislava, Slovakia}
\affiliation{Dzhelepov Laboratory of Nuclear Problems, Joint Institute for Nuclear Research, 141980~Dubna, Russia}
\author{Petr Vogel}
\email{pvogel@caltech.edu}
\affiliation{Kellogg Radiation Laboratory and Physics Department, Caltech, Pasadena, California, 91125, USA}

\begin{abstract}
The Quasiparticle Random Phase Approximation (QRPA) is used in evaluation of the total muon capture rates
for  final nuclei participating in double-beta decay. Several variants of the method are used, depending on the
size of the single particle model space used, or treatment of the initial bound muon wave function. The resulting capture rates
are all reasonably close to each other. In particular, the variant that appears to be most realistic results in rates
that are in good
agreement with the experimental values. There is no necessity for an empirical quenching of the axial current coupling
constant $g_A$. Its standard value  $g_A$ = 1.27 seems to be adequate.
\end{abstract}



\date{\today}

\maketitle

\section{Introduction}

The capture of negative muons from the $1s$ muonic atom orbit,
\begin{equation} 
\mu^- + (Z,N) \rightarrow \nu_{\mu} + (Z-1,N+1)^* ~,
\end{equation}
has been studied in detail for a long time
(see the classic reviews by  Walecka \cite{Wal75}, Mukhopadhyay \cite{Muk77}, and Measday \cite{Mea01}).  
 Experimental determination of the total muon capture rate is relatively straightforward, therefore
 it is known for many stable elements, sometimes even for the separated isotopes \cite{Eck66,Suz87}.
 
 The nuclear response in this semileptonic  weak process is governed by the momentum transfer
of the order of muon mass. 
The region of the excited nuclear states near the giant dipole resonance dominates in the
final nuclei since the phase space as well as nuclear response give preference to low excitation energies. 
These features lead to the recent revival of interest in the muon capture as a testing ground for theoretical description of weak nuclear
 processes. In particular, the question of the so-called ``axial current quenching" phenomenon is widely discussed in connection with the
 evaluation of $0\nu\beta\beta$ decay nuclear matrix elements.
 
  It is well known that using the nuclear shell model leads to
 the prediction of the allowed Gamow-Teller $\beta$ decays, as well as of the two-neutrino double beta 
 decays ($2\nu\beta\beta$), that are too fast compared to the experimental lifetimes.  
 The corresponding enhancement factors are approximately the same
 for all nuclei in the same shell, thus they can be conveniently described by  a phenomenological 
 effective axial vector coupling constant $g_A^{\rm eff}$  that is smaller than the $g_A = 1.27$ deduced from the 
 free neutron $\beta$ decay \cite{Bro88, Mar96}.
 Since the $2\nu\beta\beta$ decay is nothing else than two GT transitions occurring at once, similar quenching
appears when the $2\nu\beta\beta$ rate is calculated in the shell model \cite{Cau12}.
Recent careful analysis \cite{Gys19} suggests that when all nuclear correlations, including the effects of the
two-body weak currents, and a proper treatment of effective operators, are included, the GT transition strength
is correctly described without the need to use the quenching idea. However, the advanced treatment 
of nuclear correlations, as in Ref. \cite{Gys19}, is not yet available for evaluation of the rate of the $0\nu\beta\beta$
and $2\nu\beta\beta$ decays, or of the muon capture.

 The magnitude of quenching, i.e. the amount $q \le 1$ of the ratio  $q = g_A^{\rm eff}/g_A$,
is nuclear model dependent. For example, when the Interaction Boson Model version IBM-2 is used \cite{Bar13}, the
corresponding $q$ is considerably smaller than in the case of the shell model treatment \cite{Cau12}.
Within the Quasiparticle Random Phase Approximation (QRPA) the situation is more complex, since the
quenching amount $q$ is strongly correlated with the particle-particle effective coupling parameter $g_{pp}$
which is often adjusted to correctly describe the $2\nu\beta\beta$ decay half-life \cite{Rod03,Sim08}. 

The quenching phenomenon has been firmly established for the low momentum transfer GT-type nuclear
transitions, governed by the selection rules $\Delta I \le 1, ~\Delta \pi$ = 0, and involving dominantly 
the $\sigma \tau$ operator.
However, the neutrinoless double beta decay ($0\nu\beta\beta$) involves momentum transfer $q \sim$ 100 MeV,
with no restriction on angular momentum and parity change. This makes the muon capture, with analogous 
unrestrictive selection rules and a magnitude of the momentum transfer,
an attractive  testing ground for nuclear model description of the nuclear matrix elements
for the $0\nu\beta\beta$ decay. One of the examples of the recent effort along these lines is in Ref. \cite{Zin19} 
dedicated mostly to the experimental study of the nuclear $\gamma$ radiation following 
muon capture in the $(\mu^- , \nu~ xn)$ reactions
on the final nuclei involved in $\beta\beta$ decay. 
 
 In this work we use the QRPA to evaluate the total muon capture rates for the $0\nu\beta\beta$ decay candidate
 nuclei and compare them with the experiment. The excitation energy and multipolarity distributions are also presented.
 Previous analogous calculations of the total muon capture rate lead to ambiguous conclusions. 
 Refs. \cite{Zin06,Kol00,Mar09} use a version of
  QRPA and conclude that
 none, or only mild, quenching is needed. Similarly, the shell model applied to the muon capture on $^{16}$O in Ref. \cite{Suz18}
 also required only minimal quenching $g_A^{\rm eff}/g_A \sim 0.95$. On the other hand, in Ref. \cite{Jok19} based on QRPA quite strong
 quenching $g_A^{\rm eff}/g_A \sim 0.5$ is required to describe the total muon capture rates of the $0\nu\beta\beta$ decay
 candidate nuclei. 
 
 This motivates us to use the formalism closely related to the one used previously for the evaluation of the $0\nu\beta\beta$
 nuclear matrix elements \cite{Sim08}. The paper is organized as follows: In Section II the
 formalism is briefly described. In Section III  we discuss the choice of input parameters and
 study the corresponding uncertainties. In Section IV the results are shown and conclusion about the amount of needed $g_A$
 quenching is discussed. Our conclusions are presented in Section V.

\section{Formalism}

\subsection{Effective weak Hamiltonian and T-matrix}

The effective weak lepton-nucleus interaction Hamiltonian is of the standard form
\begin{equation}
  {H}_{\rm w}(x)=\frac{G_{\beta}}{\sqrt{2}} \bar{\nu}_{\mu}(x)\gamma_\alpha (1-\gamma_5) \mu (x)
J_L^\alpha (x) + {h.c.}.
\label{eq.2} 
\end{equation}
Here $G_\beta = G_F \cos{\theta_C}$, ${\theta_C}$ is the Cabbibo angle. 
$\mu(x)$ and $\nu_{\mu}(x)$ are the muon and muonic neutrino fields, respectively. 
$J^\alpha_L(x)$ is the V-A hadronic current at the nucleon level renormalized by
the strong and electromagnetic interactions. We have 
\begin{eqnarray}
  J_L^{\alpha} &=& \langle n(p')| \overline{d} \gamma^\alpha (1-\gamma_5) u |p(p)\rangle \nonumber\\
&=& \overline{n}(p')\left[ g_V(q^2) \gamma^\alpha + i g_M(q^2) \frac{\sigma^{\alpha\beta}}{2 m_p} q_\beta \right.\nonumber\\
    &&~~~\left. -g_A(q^2) \gamma^\alpha\gamma_5 - g_P (q^2) q^\alpha \gamma_5\right] p(p), 
\end{eqnarray} 
where $m_p$ is the nucleon mass, $q_\mu = (p'-p)_\mu$ is the momentum transfer and 
$p'$ and $p$ are the four momenta of neutron and proton, respectively. For the nucleon form factors
$g_V(q^2)$, $g_M(q^2)$, $g_A(q^2)$, and $g_P(q^2)$ we use the usual dipole parametrization
\begin{eqnarray}
  \frac{g_{V,M}(q^2)}{g_{V,M}} &=& \left(1 - \frac{q^2}{M_V^2}\right)^{-2} ~,
  \nonumber\\
  \frac{g_{A}(q^2)}{g_{A}} &=&   \left(1 - \frac{q^2}{M_A^2}\right)^{-2}~,  
\end{eqnarray} 
with $M_V = 850$ MeV and $M_V = 1086$ MeV and $g_V\equiv g_V(0)=1$, $g_A\equiv g_A(0)=1.269$,
$g_M\equiv g_M(0) = (\mu_p-\mu_n)g_V$ = 3.70. The induced pseudoscalar form factor is given by the PCAC relation
\begin{eqnarray}
  g_{P}(q^2) = \frac{2 m_p}{m^2_\pi-q^2} g_A(q^2)~,
\end{eqnarray} 
where $m_\pi$ is the pion mass.

Next, it is necessary to reduce the nucleon current to the nonrelativistic
form. By keeping terms up to $1/m_p$ and neglecting
terms ${\cal O}(q_0^2/m^2_p)$ we get \cite{EriWei} 
\begin{eqnarray}
  J^0_L &=& g_V(q^2) - g_A(q^2) \frac{\bm{\sigma}\cdot (\mathbf{p}+\mathbf{p'})}{2 m_p}
+ g_P(q^2) \frac{q_0~\bm{\sigma}\cdot\mathbf{q}}{2 m_p},\nonumber\\
&&  \nonumber\\
  \mathbf{J}_L &=& - g_A(q^2) \bm{\sigma}
  + g_P(q^2) \frac{\mathbf{q}~\bm{\sigma}\cdot\mathbf{q}}{2 m_p}
  + g_V(q^2)  \frac{\mathbf{p}+\mathbf{p'}}{2 m_p}\nonumber\\
&&  \nonumber\\
&&  + \left(g_V(q^2) + g_M(q^2)\right)~   i \frac{\bm{\sigma}\times\mathbf{q}}{2 m_p}. 
\end{eqnarray}
Note that usually non-relativistic reduction is performed in the Breit frame ($q_0=0$ and
$\mathbf{p}+\mathbf{p'}=0$ \cite{EriWei}), e.g., in the case of the $0\nu\beta\beta$-decay
\cite{Sim08} and the elastic electron nucleon (nucleus) scattering of neutrinos on nuclei, etc.
In these processes the energies of incoming and outgoing leptons  are approximately the same
or negligible.

Unlike that, the calculation of muon capture is performed in the proton rest frame where $q_0 = E_\mu - E_\nu $, 
$\mathbf{q} ~=~ \mathbf{p'}-\mathbf{p} = -\mathbf{p}_\nu$ and $\mathbf{p'}+\mathbf{p} ~= -\mathbf{p}_\nu$
\cite{Com73,Ber01}, since $p = 0$ in this frame. 
$E_\mu = m_\mu - \varepsilon_b$ is the energy of the bound muon in the $\kappa = -1$ state in the muonic atom, where
$\varepsilon_b$ is the binding energy. $E_\nu$ and $\mathbf{p}_\nu$ are energy and momentum of
emitted neutrino, respectively. $p_\nu=|\mathbf{p}_\nu|=E_\nu$ since we neglect neutrino mass.
Thus, within the non-relativistic impulse approximation, the hadronic current for muon capture
on nuclei is expressed as
\begin{eqnarray}
  J^0_L &=& g_V(q^2) + g_A(q^2) \frac{\bm{\sigma}\cdot\mathbf{p}_\nu}{2 m_p}
- g_P(q^2) \frac{q_0~\bm{\sigma}\cdot\mathbf{p}_\nu}{2 m_p},\nonumber\\
\nonumber\\
\mathbf{J}_L &=&   -g_A(q^2) \bm{\sigma} + g_P(q^2) \frac{\mathbf{p}_\nu~\bm{\sigma}\cdot\mathbf{p}_\nu}{2 m_p}
  - g_V(q^2)  \frac{\mathbf{p}_\nu}{2 m_p} \nonumber\\
&&  - i \left(g_V(q^2)+g_M(q^2)\right)  \frac{\bm{\sigma}\times\mathbf{p}_\nu}{2 m_p}. 
\end{eqnarray} 
Apart of few small terms the structure of the current is the same as in the case of the $0\nu\beta\beta$-decay. 

The muon capture on nuclei occurs in the first order in weak interaction.
The corresponding S-matrix is 
\begin{eqnarray}
  \langle f| S^{(1)}| i\rangle = 2\pi \delta(E_f + E_\nu - E_i - E_\mu)
  \langle f| T^{(1)}| i\rangle 
\end{eqnarray}  
where the T-matrix is 
\begin{eqnarray}
\langle f| T^{(1)}| i\rangle &=& (-i) \frac{G_{\beta}}{\sqrt{2}}
\int \langle f| J_L^\alpha (0,\mathbf{r}) |i\rangle \times \nonumber\\
&&  ~
\overline{\Phi}(E_\mu,\mathbf{r}) \gamma_\alpha (1-\gamma_5)\Phi(E_\nu,\mathbf{r})
~ d\mathbf{r}.
\label{Tmatrix}
\end{eqnarray}
Nuclear current takes the form 
\begin{eqnarray}
  J^\alpha_L(0,\mathbf{r}) = \sum_{n=1}^A \tau^-_n \left( g^{\alpha 0} J^0_L + g^{\alpha k} (\mathbf{J}_L)^k\right)
  \delta(\mathbf{r}-\mathbf{r}_n) 
\end{eqnarray}  
and wave functions of the bound $\kappa = -1$ muon $\Phi_{\mu}(E_\mu,\mathbf{r})$
and emitted neutrino $\Phi_{\nu}(E_\nu,\mathbf{r})$ are given by 
\begin{eqnarray} {\label{eq:muon}}
\Phi_{\mu}(E_\mu,{\mathbf r})  &=&  \frac{1}{\sqrt{4 \pi}}
\left( \begin{array}{c}
g_{-1} (r)~ \chi_{m}  \\ 
-i f_{-1} (r)~ (\bm{\sigma}\cdot \hat{\mathbf r})~ \chi_{m} 
\end{array}
\right), \nonumber\\
\Phi_{\nu}(E_\nu,{\mathbf r})  &=&  \frac{1}{\sqrt{2}}
\left( \begin{array}{c}
  ~ \chi_{m}  \\ 
\frac{\bm{\sigma}\cdot \hat{\mathbf p}_\nu}{E_\nu}~ \chi_{m} 
\end{array}
\right)~ e^{- i \mathbf{p}_\nu\cdot\mathbf{r}}. \label{wavef}
\nonumber\\ 
\end{eqnarray}
The energy of emitted neutrino follows from the energy conservation guaranteed by the $\delta$ function
and is determined by the equation
\begin{equation}
E_\nu + \sqrt{M_f^2 + p_\nu^2} - (m_\mu - \varepsilon_b + M_i) = 0. 
\end{equation}  
Here, the energies of the initial $|i\rangle$ and final $|f\rangle$ 
states are $E_i = M_i$ and $E_f = \sqrt{M_f^2 + p_\nu^2}$, respectively.  
For medium-heavy nuclei the nuclear recoil energy $p^2_\nu/(2 M_f)$ is of the order of
tens of keV and can be safely neglected, as well as the effect of center of mass
of muon-nuclear system.

\subsection{Muon capture rate}

The differential muon capture rate summed over all final excited states $|{\rm k}\rangle$
can be written as
\begin{eqnarray}
  d\Gamma = 2\pi \sum_{k} \delta(E_\nu+E_{k}-E_i-E_\mu)
\sum_{\rm spin} |\langle k|T|i\rangle|^2 
 \frac{d\mathbf{k} } {(2\pi)^3}. \nonumber\\  
\end{eqnarray}
Here, the squared T-matrix is summed over all spin orientations of the neutrino and daughter
nucleus and averaged over all spin orientations of the muon and the parent nucleus.

Inserting Eqs.  (\ref{Tmatrix}) and (\ref{wavef}) for the squared T-matrix element,
we find the total capture rate.
When only parity even operators, relevant for the ground state expectation value are kept, the total capture rate takes the form
\begin{eqnarray}
 \Gamma =   m_\mu ~\frac{\left( G_\beta m^2_\mu\right)^2}{2 \pi} \
\left(C_V B_{\Phi V} + C_A B_{\phi A} + C_P B_{\phi P}\right).\nonumber\\
  \label{eq:gamma}
\end{eqnarray}
The quantities $B_{\Phi K}$ ($K=V,A,P)$ are
\begin{eqnarray}
  B_{\Phi K} = \sum_{k} ~\frac{E_{\nu_k}^2}{m^2_\mu}~ B^k_{\Phi K}(p_{\nu_k}),
\end{eqnarray}  
where $E_{\nu_k} = p_{\nu_k} = E_\mu + E_i - E_{k}$.
The sum is over all states $|k\rangle$ in the nucleus (Z-1,N+1) that can be reached by
the corresponding operators  involved in the squared matrix elements
\begin{eqnarray}
B_{\Phi K}^k (p_{\nu_k}) &=& \frac{1}{\hat{J}_i} \sum_{M_i M_k} \int \frac{d\Omega_\nu}{4 \pi}\times\\
  && |\langle J_kM_k| \sum^A_{j=1}\tau^-_j e^{i \mathbf{p}_{\nu_k}\cdot\mathbf{r}_i} {\cal O}_K \frac{\Phi_g(r_i)}{m^{3/2}_\mu} |J_iM_i\rangle|^2.\nonumber
\end{eqnarray}  
Here, $|J_i M_i\rangle$ ($|J_k M_k\rangle$) is the initial (final) nuclear
state with spin $J_i$ ($J_k$) and spin-projection $M_i$ ($M_k$), 
$\hat{J}_i= 2J_i+1$, $\Phi_g(r)= g_{-1}(r)/\sqrt{4\pi}$ and
\begin{eqnarray}
{\cal O}_V = 1,~~~{\cal O}_A = \bm{\sigma}_j,~~~{\cal O}_P = \bm{\sigma}_j\cdot\hat{\mathbf{p}}_{\nu_k}. 
\end{eqnarray}

The effective coupling constants  in Eq. (\ref{eq:gamma}) are 
\begin{eqnarray}\label{presentc}
  C_V &=&  g_V^2(q^2) \left( 1 + \frac{p^2_\nu}{(2 m_p)^2}\right)\nonumber\\
  \nonumber\\
  C_A &=&  g^2_A(q^2) + (g_V(q^2) + g_M(q^2))^2 \frac{p^2_\nu}{(2 m_p)^2} \nonumber\\
  \nonumber\\
  C_P &=& \frac{p^2_\nu}{(2 m_p)^2}~ \left(  g^2_A(q^2) - 2 g_A(q^2) g^\mu_P(q^2)~ \frac{2 m_p}{m_\mu}
  \right. \nonumber\\
  &&\left. + (g_P^\mu)^2(q^2) \frac{p^2_\nu}{m^2_\mu} - (g_V(q^2) + g_M(q^2))^2\right). 
\end{eqnarray}  
Here, the dimensionless pseudoscalar form factor is $g^\mu_{P}(q^2) = m_\mu g_{P}(q^2) = \frac{2 m_p m_\mu}{m^2_\pi-q^2} g_A(q^2)$.
The coefficients $C_{V,A,P}$ only weakly depend on the neutrino momentum $p_{\nu_k}$.
For a sake of simplicity they are not included in the calculation of $B_{\Phi K}$ 
but are evaluated for some average neutrino momentum $p_\nu$.

Often used \cite{fuji76, Lod67, Can77, Kim85} alternative reduction of the nucleon current to its nonrelativistic form is based 
on a renormalization procedure of nucleon current due to strong interaction of Ref. \cite{Ros54}.
This, so-called Fujii-Primakoff form, of the constants $C_V$, $C_A$ and $C_P$ in Eq.
(\ref{eq:gamma}) is
\begin{equation}
\label{fpcoef}
  C_V = G_V^2,~~~C_A = G_A^2, ~~~C_P = G_P^2 - 2 G_A G_P  ~,
\end{equation}
with 
\begin{eqnarray}
  G_V &=&  g_V(q^2) \left( 1 + \frac{p_\nu}{2 m_p} \right), \nonumber\\
  \nonumber\\
  G_A &=& - g_A(q^2) - \left(g_V(q^2) + g_M(q^2) \right) \frac{p_\nu}{2 m_p}, \nonumber\\ 
  \nonumber\\
  G_P &=& \left(g^\mu_P(q^2) + g_A(q^2) - g_V(q^2) + g_M(q^2)\right) \frac{p_\nu}{2 m_p}. \nonumber\\ 
  \nonumber\\  
\end{eqnarray}  
These two forms of the constants $C_V$, $C_A$ and $C_P$ differ in the recoil order terms $p_{\nu}/(2m_p)$.

In order to make the analogy to the evaluation of the $0\nu\beta\beta$ matrix element more explicit we
introduce Fermi, Gamow-Teller and tensor squared matrix elements,
\begin{eqnarray}
B_{\Phi F} = B_{\Phi V},~~~B_{\Phi GT} = B_{\Phi A},~~~ B_{\Phi T} = 3 B_{\Phi P}- B_{\Phi A},\nonumber\\
\end{eqnarray}
Their explicit form will be given in subsection F. The muon capture rate then takes the form
\begin{eqnarray}{\label{eq:rate1}}
&&  \Gamma =   m_\mu ~\frac{\left( G_\beta m^2_\mu \right)^2}{2 \pi}  \times\\
&&~~~~\left(g_A^{\rm eff}\right)^2 \left( C_F \frac{ B_{\Phi F}}{(g^{\rm eff}_A)^2}
  + C_{GT} B_{\Phi GT} + C_T B_{\Phi T} \right),
\nonumber 
\end{eqnarray}  
where the  $(g_A^{\rm eff})^2$ appears as a scale parameter. The constants $C_F, C_T$, and $C_{GT}$
are given by
\begin{eqnarray}
  C_F &=& G^2_V, ~~~  C_T = \frac{1}{(g^{\rm eff}_A)^2}~\frac{C_P}{3},\nonumber\\
  C_{GT} &=& \frac{1}{(g^{\rm eff}_A)^2}~\left(C_A + \frac{C_P}{3} \right).
 \label{ccoef} 
\end{eqnarray}  

In Table \ref{tab.coef2} we compare the $C_F,C_{GT}$ and $C_T$ coefficients introduced
in the present approach with those governing the Fujii-Primakoff approach. We note that 
the coefficients $C_K$ ($K=F,GT,T$) are less dependent on the neutrino energy $E_{\nu}$ than
the coefficients $C_V, C_A ~{\rm and}~C_P$, and only slightly dependent on the parameter $g_A^{\rm eff}$.

\begin{table}[htb]
  \begin{center}
    \caption{\label{tab.coef2} The coefficients $C_F$, $C_{GT}$ and $C_T$ 
    (see Eq. (\ref{ccoef})) calculated
      within the present approach (see Eq.(\ref{presentc}))
      and in the Fujii-Primakoff approximation (see Eq. (\ref{fpcoef})).    }
\begin{tabular}{lcccccccccc}
  \hline\hline
$E_{\nu}$  &               & & \multicolumn{3}{c}{ present approach} & &
  \multicolumn{3}{c}{ Fujii-Primakoff} & \\ \cline{4-6}\cline{8-10}
[MeV]   & $g_A^{\rm eff}$ & & $C_F$ & $C_{GT}$ & $C_T$ & & $C_F$ & $C_{GT}$ & $C_T$ & \\\hline
  75     &   0.80        & & 0.976 & 0.797 & -0.241  & & 1.054 & 1.165 & -0.333 & \\
         &   1.00        & & 0.976 & 0.821 & -0.197  & & 1.054 & 1.091 & -0.296 & \\
         &   1.27        & & 0.976 & 0.847 & -0.158  & & 1.054 & 1.030 & -0.265 & \\
  85     &   0.80        & & 0.965 & 0.805 & -0.239  & & 1.052 & 1.203 & -0.359 & \\
         &   1.00        & & 0.965 & 0.823 & -0.197  & & 1.052 & 1.117 & -0.317 & \\
         &   1.27        & & 0.965 & 0.844 & -0.159  & & 1.052 & 1.048 & -0.282 & \\
  95     &   0.80        & & 0.955 & 0.818 & -0.234  & & 1.051 & 1.241 & -0.385 & \\
         &   1.00        & & 0.955 & 0.828 & -0.195  & & 1.051 & 1.145 & -0.337 & \\
         &   1.27        & & 0.955 & 0.844 & -0.159  & & 1.051 & 1.067 & -0.298 & \\ 
\hline\hline
\end{tabular}
  \end{center}
\end{table}

In this paper we use two alternative ways to include the bound muon wave function in the muon capture rate formula.
Traditionally, in order to simplify the calculation, it is  assumed that the muon wave function
and nuclear matrix elements can be separated. This is done by averaging muonic wave function
over the nuclear charge density distributions (to be determined and discussed later). We have then
\begin{eqnarray}
  B_{\Phi K} = \frac{\langle\Phi_{\mu}^2\rangle}{m^3_\mu}~B_K.   
\label{mwfact}
\end{eqnarray}  
Thus, while the $B_{\Phi K}$ depends on the bound muon wave function $\Phi$, in the case of the factorization
that dependence is separated into the factor $\frac{\langle\Phi_{\mu}^2\rangle}{m^3_\mu}$. The quantities $B_K$ thus
are pure nuclear quantities, independent of the muon wave function $\Phi$.

However, as we explain further  in subsection \ref{g-incl}, it is possible to include the bound muon 
wave function $g_{-1}(r)$ directly, without 
factorization. We will show later that these two alternatives lead to essentially equivalent resulting capture rates.
The final numerical results on the muon capture rate are therefore presented  with both the non-factorization and
the traditional way with factorization.  Comparison will be presented with the properly normalized
$B_{\Phi K}$:
\begin{eqnarray}{\label{eq:tilde}}
  {\widetilde B}_K  = \frac{m^3_\mu}{\langle\Phi_{\mu}^2\rangle}~B_{\Phi K}.
\end{eqnarray}

\subsection{Integration with the muon wave function}
\label{g-incl}

In all previous theoretical evaluation of the muon capture rate
the factorization in Eq. (\ref{mwfact}) was used:
\begin{eqnarray}{\label{eq:rate2}}
&&  \Gamma =   m_\mu ~\frac{\left( G_\beta m^2_\mu \right)^2}{2 \pi}
  \frac{\langle{\Phi}^2_\mu\rangle}{m^3_\mu}\times\\
&&~~~~\left(g_A^{\rm eff}\right)^2 \left( C_F \frac{{B}_F}{(g^{\rm eff}_A)^2}
  + C_{GT} {B}_{GT} + C_T {B}_T \right).
\nonumber 
\end{eqnarray}  
However,
if the quatities $B_{K}$ are replaced by ${\widetilde B}_K$ ($K = F, GT ~{\rm and}~ T$) the capture
rate without factorization of muon wave function is obtained.

The squared nuclear matrix elements ${B}_{F,GT,T}$ in Eq. (\ref{eq:rate2}) can be expressed
in general as
\begin{eqnarray}
{B}_{F} &=& \langle 0^+_i | \sum_{jk} \tau^-_j \tau^+_k
{\cal F}^{F}(\mathbf{r_j},\mathbf{r_k})|0^+_i\rangle,\nonumber\\  
{B}_{GT,T} &=& \langle 0^+_i | \sum_{jk} \tau^-_j \tau^+_k
{\cal F}^{GT,T}(\mathbf{r_j},\mathbf{r_k},\bm{\sigma_j},\bm{\sigma_k})
|0^+_i\rangle\ .\nonumber\\  
\end{eqnarray}  

Let us define the distribution functions $D_{K}(r_1)$, 
and $D_{K}(r_1,r_2)$  ($K = F, GT~ {\rm and}~ T$) as 
\begin{eqnarray}
&&{D}_{F}(r_1) = \langle 0^+_i| \sum_{jk} \tau^-_j \tau^+_k \delta(r_1-r_j)
{\cal F}^{F}(\mathbf{r_j},\mathbf{r_k})|0^+_i\rangle\nonumber\\  
&&{D}_{GT,T}(r_1) =\nonumber\\
&& \langle 0^+_i | \sum_{jk} \tau^-_j \tau^+_k \delta(r_1-r_j)
{\cal F}^{GT,T}(\mathbf{r_j},\mathbf{r_k},\bm{\sigma_j},\bm{\sigma_k})
|0^+_i\rangle ~, \nonumber\\  
\end{eqnarray}  
as well as
\begin{eqnarray}
&&{D}_{F}(r_1,r_2) = \nonumber\\
&&  \langle 0^+_i| \sum_{jk} \tau^-_j \tau^+_k \delta(r_1-r_j)\delta(r_2-r_k)
{\cal F}^{F}(\mathbf{r_j},\mathbf{r_k})|0^+_i\rangle\nonumber\\  
&&{D}_{GT,T}(r_1,r_2) =\langle 0^+_i | \sum_{jk} \tau^-_j \tau^+_k 
\delta(r_1-r_j) \delta(r_2-r_k)\nonumber\\
&&~~~~~~~~~~~~~~~~~~~~~~~~~~
{\cal F}^{GT,T}(\mathbf{r_j},\mathbf{r_k},\bm{\sigma_j},\bm{\sigma_k})|0^+_i\rangle  
\end{eqnarray}

Obviously, the $D$-functions are normalized as
 \begin{eqnarray}
&&{B}_{K} =  \int_0^\infty D_{K}(r_1) d r_1, \nonumber \\
&& {B}_{K} =  \int_0^\infty D_{K}(r_1,r_2) d r_1 d r_2.
\end{eqnarray}

Once the $D_K(r_1,r_2)$ have been calculated one can avoid the factorization of
the (averaged) muon wave function and of the sum of
squared nuclear matrix elements. The relevant quantity is then
\begin{eqnarray}
\frac{1}{4\pi} \frac{1}{m^3_\mu} \int_0^\infty g_{-1}(r_1) g_{-1}(r_2)D_{K}(r_1,r_2) d r_1 d r_2
\end{eqnarray}
instead of ${{\Phi}^2_\mu}/{m^3_\mu}~ B_{K}$ in the Eq.(\ref{eq:rate1}).
We will show below that the two alternative approaches of treating the bound muon wave function lead
to very similar muon capture rates.

\subsection{Separation of the muon wave function}

For medium and heavy nuclei considered in the present work the relativistic effects on the bound muon are essential, thus
the muon wave function is obtained by solving the Dirac equation.  In it the nuclear potential is based on the Fermi type
charge distribution, with parameters specified in Table \ref{tab:zeff}.

The wave function of the bound $\kappa=-1$ muon is given in Eq. (\ref{eq:muon}).
The effect of the nuclear charge distribution on the muon, relevant for the muon capture, can
be described by the overlap
\begin{equation}
  \langle {\Phi}^2_\mu\rangle_g = 
    \int_0^\infty \frac{g^2_{-1}(r)}{4\pi} \rho(r) r^2 dr 
\end{equation}
or, essentially equivalently
\begin{equation}
     \langle {\Phi}^2_\mu\rangle_Z = 
    \int_0^\infty \frac{(g^2_{-1}(r)+f_{-1}^2(r))}{4\pi} \rho(r) r^2 dr  ~.
\end{equation} 
The nuclear charge distribution $\rho(r) \sim 1/(1 + exp((r - c_{\rm rms})/a))$ is normalized to $Z$, the proton number.

\begin{center}
  \begin{figure}[htb]
  \includegraphics[width=0.50\textwidth]{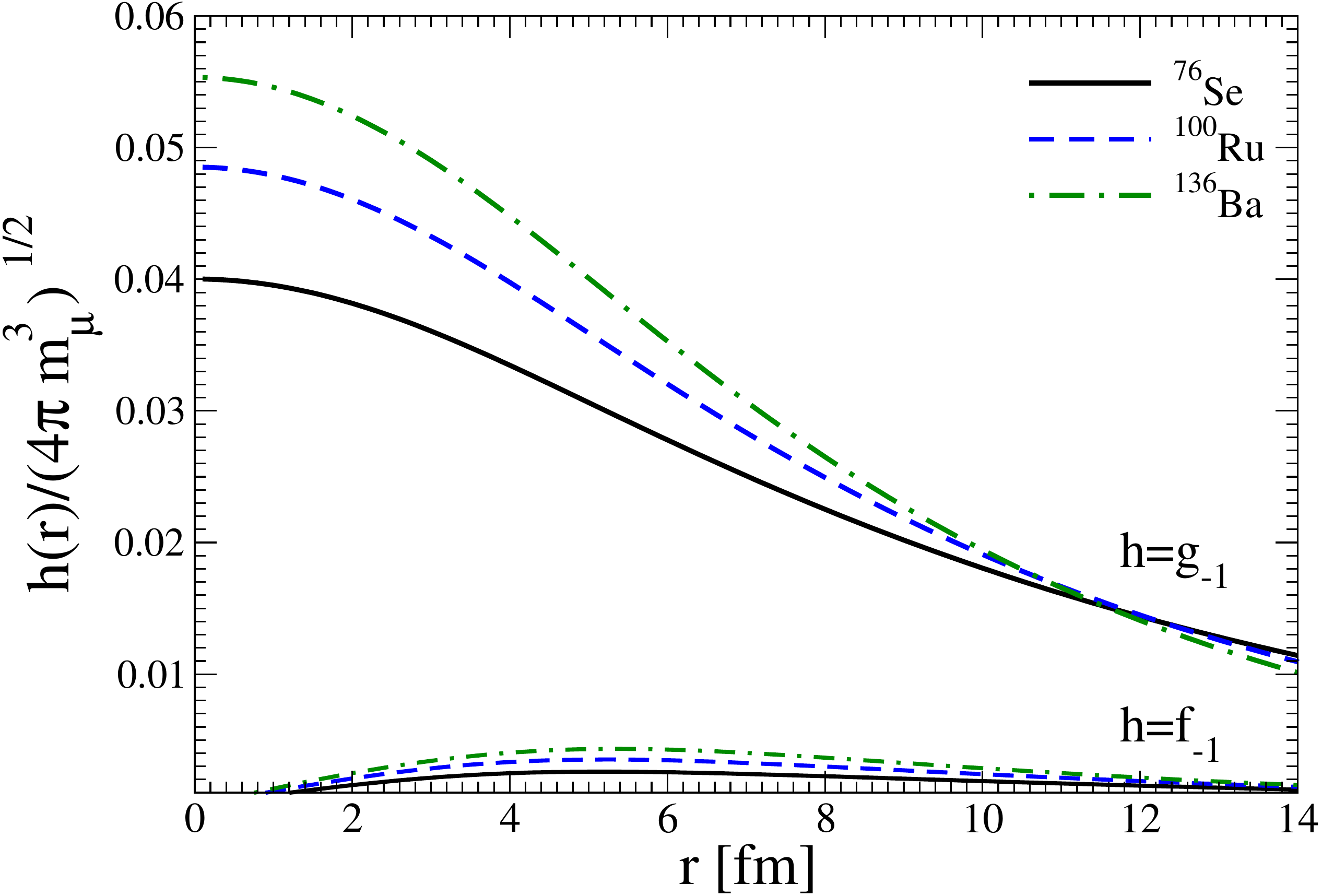}
  \caption{(Color online) The radial dependence of the bound muon wave functions $g_{-1}(r) $ and 
   $f_{-1}(r) $ for $^{76}$Se, $^{100}$Ru and $^{136}$Ba. }
\label{fi:gwfr}
\end{figure}
\end{center}

In Fig. \ref{fi:gwfr} we show examples of the radial muon wave functions $g_{-1}(r)$ and  $f_{-1}(r)$. Since the small
component $f_{-1}(r)$
vanishes at the origin its effect on the muon capture is negligible. 

Traditionally \cite{For62} the quantity $\langle {\Phi}^2_\mu\rangle$ is replaced by the
empirical parameter $Z_{eff}$ using
\begin{eqnarray}
  Z^4_{\rm eff}  &=& \frac{\pi~Z}{\alpha^3}~\frac{\langle {\Phi}^2_\mu\rangle_{_Z}}{m^3_\mu},\nonumber\\  
  \frac{\langle {\Phi}^2_\mu\rangle_{_Z}}{m^3_\mu} &=& \frac{\alpha^3}{\pi~Z}~Z^4_{\rm eff}.
\end{eqnarray}
Using $\langle {\Phi}^2_\mu\rangle_g$ instead of $\langle {\Phi}^2_\mu\rangle_Z$ makes little difference
as seen in Table \ref{tab:zeff}. Also, we tested that $Z_{eff}^4$ is insensitive to variations of $c_{\rm rms}$.
By changing it by 1\%  changes $Z_{eff}^4$ also by approximately the same amount. Replacing the Fermi
distribution by the sharp surface nuclear charge distribution also changes $Z_{eff}^4$ by only a small amount.

\begin{table*}[htb]
  \begin{center}
    \caption{\label{tab:zeff} The effective charge $Z_{\rm eff}$ of Ref. \cite{Suz18} is shown in col. 2
    and $Z_{\rm eff}$ determined in the present work in col. 9. The half-way radius $c_{\rm rms}$ is in col. 6. The Fermi
     distribution $\rho(r) = 1/(1 + exp((r-c_{\rm rms})/a))$ is chosen such that the mean square radius
     $\langle r^2 \rangle$ has its experimental value \cite{Ang13} shown in col. 5; the surface thickness $a$ = 0.545 fm
     is used.     The muon wave function $\langle {\Phi}^2_\mu\rangle$/$m_{\mu}^3$
     averaged over the nuclear charge distribution  $\rho(r)$ is shown in col. 7  using $g_{-1}^2(r) + f_1^2(r)$ and only
      $g_{-1}^2(r)$ in col. 8.  $\varepsilon_b$ in col 4 is the binding energy of the muon in the $\kappa = -1$ state.}
\begin{tabular}{lccccccccccc}
  \hline\hline
Elem. & $Z_{\rm eff}$ & & Nucl. & $\varepsilon_b$ & $\sqrt{\langle r^2 \rangle}$ & $c_{\rm rms}$ & &
\multicolumn{3}{c}{$~~\rho(r) = \rho_0 (1+exp((r-c_{\rm rms})/a))^{-1}$}  \\ \cline{9-11}
        &   Ref. \cite{Suz18} & & & [MeV]  & [fm]  & [fm]   &  &  
$\langle {\Phi}^2_\mu\rangle_{g}/m^3_\mu$ &  $\langle {\Phi}^2_\mu\rangle_{_Z}/m^3_\mu$ & $Z_{\rm eff}$ & \\ \hline
Ti &  17.38 & & ${^{48}_{22}{\rm Ti}}$ & 1.268  &  3.592 &  3.828  & & 
             $5.449\times 10^{-4}$ & $5.461\times 10^{-4}$ & 17.654 & \\
Se &  23.24 & & ${^{76}_{34}{\rm Se}}$ & 2.760  &  4.140 &  4.659  &  &
             $1.087\times 10^{-3}$ & $1.092\times 10^{-3}$ & 23.405 & \\
Kr &        & & ${^{82}_{36}{\rm Kr}}$ & 3.046  &  4.192 &  4.737  &   &
             $1.177\times 10^{-3}$ & $1.183\times 10^{-3}$ & 24.223  & \\
Mo & 26.37  & & ${^{96}_{42}{\rm Mo}}$ & 3.939  &  4.385  &  5.019  &   &
             $1.407\times 10^{-3}$ & $1.415\times 10^{-3}$ & 26.328 & \\
Ru  &       & & ${^{100}_{~44}{\rm Ru}}$ & 4.247 &  4.453 &  5.119  &   &
             $1.470\times 10^{-3}$ & $1.480\times 10^{-3}$ & 26.935 & \\
Cd  & 28.20 & & ${^{110}_{~48}{\rm Cd}}$ & 4.877 &  4.577 &  5.297   &  &
             $1.588\times 10^{-3}$ & $1.600\times 10^{-3}$ & 28.069 & \\
Sn  & 28.64 & & ${^{116}_{~50}{\rm Sn}}$ & 5.203 &  4.625 &  5.367  &  &
             $1.647\times 10^{-3}$ & $1.660\times 10^{-3}$ & 28.621 & \\
Te  & 29.03 & & ${^{124}_{~52}{\rm Te}}$ & 5.513 & 4.718 &  5.500  &  &
             $1.673\times 10^{-3}$ & $1.686\times 10^{-3}$ & 29.017 & \\
Xe &        & & ${^{128}_{~54}{\rm Xe}}$ & 5.838 & 4.777 &  5.585  &  &
             $1.715\times 10^{-3}$ & $1.729\times 10^{-3}$ & 29.477 & \\
   &        & & ${^{130}_{~54}{\rm Xe}}$ & 5.836 & 4.782 &  5.591  &  &
             $1.712\times 10^{-3}$ & $1.726\times 10^{-3}$ & 29.464 & \\
Ba & 29.99  & & ${^{134}_{~56}{\rm Ba}}$ & 6.168 & 4.832 & 5.663  &  &
             $1.755\times 10^{-3}$ & $1.771\times 10^{-3}$ & 29.922 & \\
    &       & & ${^{136}_{~56}{\rm Ba}}$ & 6.167 & 4.833 & 5.664  &  &
             $1.754\times 10^{-3}$ & $1.770\times 10^{-3}$ & 29.919 & \\
Sm  & 31.01 & & ${^{150}_{~62}{\rm Sm}}$ & 7.140 & 5.039 & 5.955  &  &
             $1.819\times 10^{-3}$ & $1.837\times 10^{-3}$ & 30.978 & \\
\hline\hline
\end{tabular}
  \end{center}
\end{table*}

\subsection{Evaluation of the matrix elements through a product of two one-body matrix elements}
\label{sub:one}

When considering muon capture on even-even nuclei with the $0^+$ ground state (all ground states relevant
for double-beta decays are $0^+$ ground states) we can simplify the evaluation of the quantities $B_{V,A,P}$ 
(without the factor $\Phi_{\mu}^2/m_{\mu}^3$) in
Eq. (\ref{eq:gamma}) to find
($I=V, A, P$)
\begin{eqnarray}
\label{bdef}
B_I  =  4\pi \sum_{J^{\pi}_k} ~\frac{E_{\nu_k}^2}{m^2_\mu} ~B^k_I(J^\pi_k, p_{\nu_k}), 
\end{eqnarray}
where $p_{\nu_k} = E_{\nu_k} = E_\mu + E_i - E_{J^\pi_k}$ and
\begin{eqnarray}
&& B^k_{V,P}(J^\pi_k, p_{\nu_k}) = \left|\sum_{p n}
\langle n\parallel {\cal O}^{V,P}_J(p_{\nu_k})\parallel p\rangle~T_{p n}(J^\pi_k)\right|^2
\nonumber\\
&& B^k_{A}(J^\pi_k, p_{\nu_k}) =  
\nonumber\\
&&~~~~~~~\sum_{L=J,J\pm1} \left|\sum_{p n}
\langle n\parallel {\cal O}^{A}_{LJ}(p_{\nu_k})\parallel p\rangle~T_{p n}(J^\pi_k)\right|^2
\end{eqnarray}
with
\begin{eqnarray}
&& {\cal O}^{V}_{JM} (p_\nu) = i^J~j_J(p_\nu r) Y_{JM}(\Omega_r),
\nonumber\\
&& {\cal O}^{A}_{LJM} (p_\nu) = i^L~ j_L(p_\nu r)
\{Y_{L}(\Omega_r)\otimes \sigma_1\}_{JM},
\nonumber\\
&& {\cal O}^{P}_{JM} (p_\nu) = i^{J-1}~\times\nonumber\\
&& ~\left( \sqrt{\frac{2J-1}{2J+1}} j_{J-1}(p_\nu r) C^{J0}_{J-1 0~10}
\{ Y_{J-1}(\Omega_r)\otimes \sigma_1 \}_{JM} \right. \nonumber\\
&& \left. - \sqrt{\frac{2J+3}{2J+1}} j_{J+1}(p_\nu r) C^{J0}_{J+1 0~10}
\{Y_{J+1}(\Omega_r)\otimes \sigma_1\}_{JM}~ \right).\nonumber\\
\end{eqnarray}
and
\begin{eqnarray}
\label{troneb}  
T_{p n}(J^\pi_k) &=& \frac{1}{\sqrt{2J+1}}~ \langle  J^{\pi}_k \parallel [c_n^+ \tilde{c}_p]_J \parallel 0_i^+ \rangle,
\nonumber\\
&=& v_p u_n X^k_{np J} +  u_p v_n Y^k_{np J}.
\end{eqnarray}
The one-body operators
$c_p^+ \tilde{c}_n$ (the tilde denotes the time-reversed state)
 appear in the reduced matrix elements. 
In them $c_n^+$ create a neutron, and $c_p$ annihilates a proton.
Such matrix elements in  Eq. (\ref{troneb})
depend on the BCS coefficients $u_\tau,v_\tau$ ($\tau = p,n$) and on the QRPA
vectors  $X^k_{np J}$ and  $Y^k_{np J}$. 
The nuclear structure information resides in these quantities.

\subsection{Calculation through  two-body matrix elements} 
\label{sub:two}

There is an alternative and equivalent way to evaluate
the squared matrix elements
$B_{F,GT,T}$ (K = Fermi (F), Gamow-Teller (GT), and Tensor (T)).
It can be expressed as sums 
over the final states, labeled by their angular momentum and parity
$J^{\pi}$ and indices $k$  in the QRPA as follows ($K=F, ~GT ~{\rm and}~ T$):
\begin{eqnarray}
\label{eq:long}
{B}_K  &=&  \sum_{J^{\pi}_k,\mathcal{J}} \sum_{pnp'n'}
\frac{E_{\nu_k}^2}{m^2_\mu} \times \\
&& (-1)^{j_n + j_{p'} + J + {\mathcal J}} ~\sqrt{ 2 {\mathcal J} + 1}
\left\{
\begin{array}{c c c}
j_p & j_n & J  \\
 j_{n'} & j_{p'} & {\mathcal J}
\end{array}
\right\}  \times\nonumber \\
&& ~\langle n(1), p'(2); {\mathcal J} \parallel 
{O}_K(p_{\nu_k}) 
\parallel p(1), n'(2); {\mathcal J} \rangle \times~~
\nonumber \\
&&~~~ \langle 0_i^+ \parallel
[ \widetilde{c_{p'}^+ \tilde{c}_{n'}}]_J \parallel J^{\pi}_k \rangle
 \langle  J^{\pi}_k \parallel [c_n^+ \tilde{c}_p]_J \parallel 0_i^+ \rangle ~.
\nonumber
\end{eqnarray}
As in the Eq. (\ref{bdef}) the reduced matrix elements of the one-body operators
$c_p^+ \tilde{c}_n$ depend on the BCS coefficients $u_i,v_j$ and on the QRPA vectors.

The two-body operators $O_{F,GT,T}$ are given by
\begin{eqnarray}
\left\{\begin{array}{c}
O_F(p_{\nu}) \\
O_{GT}(p_{\nu}) \\
O_T(p_{\nu}) 
\end{array}
\right\} = 
\tau^+_1\tau^-_2~
\left\{\begin{array}{l}
~~j_0(p_{\nu} r_{12}) \\
-j_0(p_{\nu} r_{12}) ~\sigma_{kl}\\
~~j_2(p_{\nu} r_{12})~ S_{12}
\end{array}
\right\}. 
\end{eqnarray}
Their matrix elements depend on the relative distance $r_{12}$.

In the above derivation we used 
\begin{eqnarray}
&&  \int e^{i \mathbf{p}_\nu\cdot\mathbf{r}_k} e^{- i \mathbf{p}_\nu\cdot\mathbf{r}_j} \frac{d\Omega_\nu}{4 \pi}
  = j_0(p_\nu r_{kj}),\nonumber\\
&&  \int \left(\bm{\sigma}_k\cdot\hat{\mathbf{p}}_\nu\right)
  \left(\bm{\sigma}_j\cdot\hat{\mathbf{p}}_\nu\right)
  e^{i \mathbf{p}_\nu\cdot\mathbf{r}_k} e^{- i \mathbf{p}_\nu\cdot\mathbf{r}_j} \frac{d\Omega_\nu}{4 \pi} \nonumber\\
  &&~~~~~ = \frac{1}{3}~\left[ j_0(p_\nu r_{kj})~\sigma_{kj}
~-~j_2(p_\nu r_{kj}) S_{kj}(\hat{r}_{kj}) \right]\nonumber\\
\end{eqnarray}
with 
\begin{eqnarray}
\sigma_{kj} &=&\bm{\sigma}_k\cdot\bm{\sigma}_j\nonumber\\
S_{kj}(\mathbf{r}_{kj}) &=& 3~ {\bm{\sigma}_k\cdot\mathbf{\hat{r}}_{kj}}
~{\bm{\sigma}_j\cdot\mathbf{\hat{r}}_{kj}}
-{\bm{\sigma}_k\cdot\bm{\sigma}_j}.
\end{eqnarray}

Note that the squared matrix elements $B_F$, $B_{GT}$ and $B_T$ are analogous to the matrix elements
associated with  the second order process contributing to electron scattering on nuclei. These
matrix elements contain a summation over  pairs of nucleons inside the nucleus with relative distance $r_{ij}$.

\section{Choice of input parameters and sensitivity of the results}
\label{sec:par}

In this section we discuss the choice of  empirical input parameters and the sensitivity of calculated rates to
them. 

The first choice to be made are the nuclear single particle energies and the corresponding wave functions.
The eigenvalues of the Coulomb-corrected Woods-Saxon potential with Bertsch parametrization
\cite{Ber72} are used. In order to test the dependence on the single particle basis we performed our calculation
with two choices of single nucleon basis. The small basis has 11 levels (oscillator shells N=0-3 plus the $g_{9/2}$ from N=4)
for $^{48}$Ti, 16 levels (oscillator shells N=0-4 plus the $h_{11/2}$ from N=5) for $^{76}$Se and $^{82}$Kr, 
18 levels (oscillator shells N=0-4 plus the $p_{3/2}$, $f_{7/2}$ and  $h_{11/2}$ from N=5) for $^{96}$Mo and $^{100}$Ru,
21 levels (oscillator shells N=0-5) for $^{110}$Cd, 22 levels  (oscillator shells N=0-5 plus the $i_{13/2}$ from N=6)
for $^{116}$Sn, $^{124}$Te, $^{128}$Xe, $^{130}$Xe, $^{134}$Ba, $^{134}$Ba and $^{136}$Ba, 23 levels  (oscillator shells N=0-5
plus the $g_{9/2}$ and $i_{13/2}$ from N=6) for $^{150}$Sm. All single particle states in the small basis are bound.

 The large model space contains 28 levels (oscillator shells N=0-6) for 
$^{48}$Ti, $^{76}$Se, $^{82}$Kr, $^{96}$Mo, $^{100}$Ru, $^{110}$Cd, $^{116}$Sn and 35 levels
(oscillator shells N=0-7 without $j_{13/2}$ from N=7) for $^{116}$Sn, $^{124}$Te, $^{128}$Xe, $^{130}$Xe, $^{134}$Ba, 
 $^{136}$Ba, and $^{150}$Sm.  Some of the neutron states in the large basis are quasibound or truly unbound.

Our results suggest that the smaller basis is inadequate since adding additional states changes
the capture rate significantly. To test the convergence of the larger single particle space we checked
that subtracting few upper levels makes only small difference.

In QRPA we treat the muon capture as the creation of the correlated proton hole - neutron particle states. Experimentally
only a fraction of the final states remains bound in the final odd-odd (Z-1,N+1) nucleus, while most final states involve the emission
of one or more neutrons. It is therefore clear that highly excited states in the final nucleus are present. Hence, it is important to include
in the calculation as many neutron single particle states above the Fermi level as possible.  On the other hand, the quasi bound or
unbound states included in the large single particle space in this work do not have the correct asymptotic behavior. It is,
therefore, likely that the optimal single particle space is between the boundaries developed in this work.

\begin{center}
  \begin{figure}[htb]
  \includegraphics[width=0.50\textwidth]{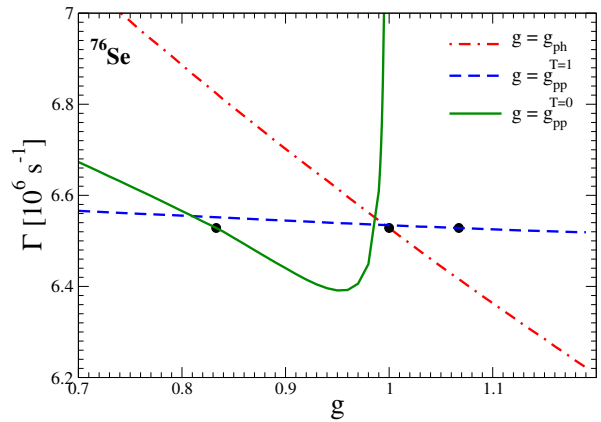}
  \caption{(Color online) The muon capture rate of $^{76}$Se as function of constants $g_{ph}$, $g^{T=1}_{pp}$ and $g^{T=0}_{pp}$  
    used to renormalize, respectively, the particle-hole, isovector and isoscalar channels of particle-particle interaction of the nuclear
    Hamiltonian. The small model space and factorization of muon wave function and nuclear matrix elements is assumed.
    The positions of the black points indicate the fixed values of these parameters in general calculation of the capture rate, namely
    $g_{ph}=1.000$, $g^{T=1}_{pp}=1.067$ and $g^{T=0}_{pp}=0.833$.  
}
\label{fi:gdep}
\end{figure}
\end{center}

The residual nuclear interaction enters
the QRPA equation of motion. In this
work we use the Brueckner G-matrix elements \cite{Mue99} calculated with a realistic
one-boson-exchange Argonne V18 potential \cite{Wir95}.

The pairing interaction has been included in the standard way, i.e. the coupling constant $g_{pair}$ of the $T=1, J=0$ interaction was slightly renormalized in order to reproduce the experimental pairing gaps. In addition,
it is customary in the application of the QRPA method in the evaluation of the $\beta\beta$ nuclear matrix elements to
adjust the particle-hole coupling parameter $g_{ph}$ as well as the two isospin components of the particle-particle coupling
parameter $g_{pp}^T$ (see Ref. \cite{Sim13,Sim18}). 

We show an example how the capture rate depends on the 
renormalized parameters $g_{ph}$ and $g_{pp}^T$ in Fig. \ref{fi:gdep}. Over the whole range
of realistic $g$ values the capture rate changes by less than 10\%, with the exception of the known singularity for $g_{pp}^{T=0} > 1$.
Our calculation use safely smaller values of  $g_{pp}^{T=0}$. For each model space we fixed values of these coupling constants
by the condition of the partial restoration of SU(4) symmetry applied for a corresponding double-beta decay transitions, as
described in Ref. \cite{Sim18}.

\begin{table}[htb]
  \begin{center}
    \caption{\label{tab:FGTT} The squared matrix elements ${B}_K$ 
     ($K = F, GT ~{\rm and} ~T$)
     for daughter isotopes of the double-beta decay transitions.
     $B_K$ and ${\widetilde B}_K$ correspond to cases with and without
     factorization of the muon wave function.
     Small (s) and large (l) single particle model spaces are considered. }
\begin{tabular}{lcccccccccc}
  \hline\hline
   &  &  &  &   &  &  &  &  &    &  \\ 
  Nucl. & m.s. &  & \multicolumn{3}{c}{$B_K$} & & \multicolumn{3}{c}{${\widetilde B}_K$} &
  \\\cline{4-6}\cline{8-10}
        &      &  &  F  & GT  &  T &  & F & GT  & T &  \\ \hline
  ${^{48}_{}{\rm Ti}}$   & s & & 2.010  & 5.715   & 0.460  &
                                 & 1.845  & 5.253   & 0.433  & \\
                         & l  &  & 2.379  & 7.308   & 1.239  &
                                 & 2.230  & 6.829   & 1.269  & \\
  ${^{76}_{}{\rm Se}}$   & s & & 3.140  & 8.697   & 1.098  &
                                 & 2.802  & 7.735   & 1.005 & \\
                         & l  &  & 3.620  & 10.37   & 1.948 &
                                 & 3.336  & 9.473   & 1.931 & \\
  ${^{82}_{}{\rm Kr}}$   & s & & 2.938  & 8.356   & 1.107 &
                                 & 2.614  & 7.385   & 1.001 & \\
                         & l  &  & 3.566  &10.303   & 2.060 &
                                 & 3.303  & 9.426   & 2.025 &  \\
  ${^{96}_{}{\rm Mo}}$   & s & & 3.514  & 9.493   & 1.098 & 
                                 & 3.171  & 8.478   & 1.008 &  \\
                         & l  &  & 4.249  &12.301   & 2.289 &
                                 & 3.908  &11.084   & 2.237 &  \\
  ${^{100}_{}{\rm Ru}}$ & s & & 3.627  & 9.175   & 1.011  &
                                 & 3.246  & 8.169   & 0.923 &  \\  
                         & l  &  & 4.485  &12.765   & 2.290 &
                                 & 4.091  &11.405   & 2.229 & \\
  ${^{110}_{}{\rm Cd}}$ & s & & 4.028  &11.593   & 1.845 &
                                 & 3.629  &10.175   & 1.669 & \\
                         & l  &  & 4.703  &13.169   & 2.426 &
                                 & 4.273  &11.706   & 2.312 &  \\
  ${^{116}_{}{\rm Sn}}$ & s & & 4.462  &11.734   & 1.631 &
                                 & 3.892  &10.100   & 1.474 &  \\
                         & l  &  & 4.733  &12.990   & 2.399 &
                                 & 4.275  &11.464   & 2.258 &  \\
  ${^{124}_{}{\rm Te}}$ & s & & 3.544  & 9.925   & 1.426 &
                                 & 3.126  & 8.627   & 1.294 &  \\
                         & l  &  & 3.966  &11.407   & 2.351 &
                                 & 3.692  &10.299   & 2.331 & \\ 
  ${^{128}_{}{\rm Xe}}$ & s & & 3.611  &10.179   & 1.462 &
                                 & 3.170  &8.818    & 1.321 &  \\
                         & l  &  & 4.193  &12.084   & 2.455 &
                                 & 3.876  &10.864   & 2.414 &  \\
  ${^{130}_{}{\rm Xe}}$ & s & & 3.277  & 9.452   & 1.415 &
                                 & 2.906  & 8.229   & 1.282 &  \\
                         & l  &  & 3.877  &11.322   & 2.380 &
                                 & 3.634  &10.251   & 2.349 & \\
  ${^{134}_{}{\rm Ba}}$ & s & & 3.373  & 9.796   & 1.432 &
                                 & 2.953  & 8.429   & 1.280 &  \\
                         & l  &  & 4.152  &12.153   & 2.472 &
                                 & 3.842  &10.891   & 2.414 &  \\
  ${^{136}_{}{\rm Ba}}$ & s & & 3.065  & 9.170   & 1.415 &
                                 & 2.704  & 7.907   & 1.265 &  \\
                         & l  &  & 3.866  &11.505   & 2.449 &
                                 & 3.617  &10.357   & 2.394 &  \\
  ${^{150}_{}{\rm Sm}}$ & s & & 3.575  &10.057   & 1.561 &
                                 & 3.247  & 8.964   & 1.427 & \\
                         & l  &  & 4.627  &13.317   & 2.803 &
                                 & 4.383  &12.191   & 2.772 &  \\
\hline\hline
\end{tabular}
  \end{center}
\end{table}

In order to compare the results with and without factorization of the muon
wave function, as well as the results with the small and large single particle 
model spacec, we show
in Table \ref{tab:FGTT} the calculated squared nuclear matrix elements 
${B}_K$ and ${\widetilde B}_ K$ (see Eq. (\ref{eq:tilde}) for the definition)
for the 13 final nuclei participating
in the double-beta decay transitions. Typically,
the Gamow-Teller matrix elements are dominant. However, the Fermi and
Tensor matrix elements give a non-negligible 
contributions. Note that the ratio $B_{GT}/B_F$ is on average about 2.8 $\pm$ 0.1, close to
the value $B_{GT}/B_F = 3$ corresponding to the pure $S= 0$ state.

The squared matrix elements using the large model space
are about 10-20\% larger in comparison with those for the small model space. 
The capture rate with and without  factorization of the muon wave function can be obtained 
by inserting  ${B}_K$ and ${\widetilde B}_K$ into Eq. (\ref{eq:rate1})
respectively. Thus, a difference of these squared matrix elements quantifies
the effect of the factorization treatment. It is not very significant, but is increasing with Z of the nucleus.
The entries weakly depend on the
$g_A^{\rm eff}$ value. They were evaluated with the $g_A^{\rm eff}$ that reproduces the empirical value of the muon capture rate.

In Table \ref{tab:mult2} we show the contributions of individual $J^{\pi}$ multipoles to the matrix elements and the total capture rate
for $^{76}$Se and $^{136}$Ba. The entries were evaluated without the factorization of the muon wave function, using the small
and large single particle model spaces. The present way of choosing the nonrelavistic reduction of the weak hamiltonian
was used. In both cases the $1^-$, $2^-$, $1^+$ and $2^+$ multipoles account for 70-80\% of the capture rate.

\begin{table*}[htb]
  \begin{center}
    \caption{\label{tab:mult2}
      The multipole decomposition of the matrix elements ${\widetilde B}_ K$ (K=V, A, P, F, GT and T)
    (see Eqs. (\ref{eq:long}) or (\ref{bdef}))
     and muon capture rate $\Gamma_{\rm pres}$ for $^{76}$Se and $^{136}$Ba  evaluated in the small (s) and large (l) single particle
      model spaces.}
\begin{tabular}{lcccccccccccccccc}
  \hline\hline
  Nucl.  &  ms & I & $0^+$ & $1^+$ & $2^+$ & $3^+$ & $4^+$ & $5^+$ & $0^-$ & $1^-$ & $2^-$ & $3^-$ & $4^-$ & $5^-$ & all &  \\ \hline
  &   &  &  &  &  &  &  &  &  &  &  &  &  &  &   &  \\  
 &   &  & \multicolumn{13}{c}{${\widetilde B}_K(J^{\pi})$} &  \\ \cline{3-16}   
${^{76}_{}{\rm Se}}$ 
  & s & V,F    & 0.306 & 0.000 & 0.561  & 0.000 & 0.008  & 0.000 &  0.000 &  1.808 & 0.000 &  0.119 & 0.000 &  0.001 & 2.802  &  \\
  &   & A, GT  & 0.000 & 1.125 & 0.439  & 0.943 & 0.007  & 0.013 &  0.983 &  1.773 & 2.156 &  0.106 & 0.187 &  0.001 & 7.735  &  \\
  &   & P      & 0.000 & 0.504 & 0.000  & 0.410 & 0.000  & 0.006 &  0.983 &  0.000 & 0.926 &  0.000 & 0.084 &  0.000 & 2.914  &  \\
  &   & T      & 0.000 & 0.387 & -0.439 & 0.286 & -0.007 & 0.005 &  1.965 & -1.773 & 0.622 & -0.105 & 0.065 & -0.001 & 1.005  &  \\

  & l & V,F    & 0.570 & 0.000 & 0.711  & 0.000 & 0.011  & 0.000 &  0.000 &  1.901 & 0.000 &  0.143 & 0.000 &  0.001 & 3.336  &  \\
  &   & A, GT  & 0.000 & 2.091 & 0.635  & 1.093 & 0.011  & 0.019 &  1.085 &  1.700 & 2.472 &  0.131 & 0.233 &  0.002 & 9.473  &  \\
  &   & P      & 0.000 & 1.012 & 0.000  & 0.481 & 0.000  & 0.009 &  1.085 &  0.000 & 1.108 &  0.000 & 0.106 &  0.000 & 3.801  &  \\
  &   & T      & 0.000 & 0.946 & -0.635 & 0.351 & -0.011 & 0.007 &  2.169 & -1.700 & 0.851 & -0.131 & 0.083 & -0.001 & 1.931  &  \\
  &   &  & \multicolumn{13}{c}{${\Gamma_{\rm pres}}(J^+)/\Gamma_{\rm pres}$} &  \\ \cline{4-16}
  & s &        & 0.019 & 0.116 & 0.091  & 0.098 & 0.001  & 0.001 &  0.071 &  0.338 & 0.224 & 0.021  & 0.019 &  0.000 & 1.000  &  \\
  & l &        & 0.035 & 0.167 & 0.110  & 0.090 & 0.002  & 0.002 &  0.059 &  0.294 & 0.201 & 0.022  & 0.019 &  0.000 & 1.000   &  \\ \hline
  &   &  &  &  &  &  &  &  &  &  &  &  &  &  &   &  \\    
        &   &  & \multicolumn{13}{c}{${\widetilde B}_K(J^{\pi)}$} &  \\ \cline{3-16}
 ${^{136}_{}{\rm Ba}}$ 
  & s & V,F    & 0.753 & 0.000 & 0.859  & 0.000 & 0.022  & 0.000 &  0.000 &  0.933 & 0.000 &  0.135 & 0.000 &  0.002 & 2.704  &  \\
  &   & A, GT  & 0.000 & 2.354 & 0.684  & 1.012 & 0.019  & 0.028 &  1.018 &  1.428 & 1.007 &  0.134 & 0.219 &  0.002 & 7.907  &  \\
  &   & P      & 0.000 & 1.045 & 0.000  & 0.448 & 0.000  & 0.017 &  1.018 &  0.000 & 0.434 &  0.000 & 0.098 &  0.000 & 3.057  &  \\
  &   & T      & 0.000 & 0.780 & -0.684 & 0.333 & -0.019 & 0.010 &  2.036 & -1.428 & 0.296 & -0.134 & 0.075 & -0.002 & 1.265  &  \\

  & l & V,F    & 1.092 & 0.000 & 1.023  & 0.000 & 0.027  & 0.000 &  0.000 &  1.277 & 0.000 &  0.195 & 0.000 &  0.002 & 3.617  &  \\
  &   & A, GT  & 0.000 & 3.465 & 0.898  & 1.154 & 0.026  & 0.038 &  1.192 &  1.530 & 1.561 &  0.183 & 0.304 &  0.002 & 10.357  &  \\
  &   & P      & 0.000 & 1.676 & 0.000  & 0.517 & 0.000  & 0.017 &  1.192 &  0.000 & 0.709 &  0.000 & 0.137 &  0.000 & 4.250 &  \\
  &   & T      & 0.000 & 1.563 & -0.898 & 0.398 & -0.026 & 0.014 &  2.385 & -1.530 & 0.565 & -0.183 & 0.106 & -0.002 & 2.394  &  \\
        &   &  & \multicolumn{13}{c}{${\Gamma}_{\rm pres}(J^+)/\Gamma_{\rm pres}$} &  \\ \cline{4-16}
  & s &        & 0.048 & 0.239 & 0.141  & 0.103 & 0.004  & 0.003 &  0.072 &  0.240 & 0.103 & 0.025  & 0.022 &  0.000 & 1.000  &  \\
  & l &        & 0.070 & 0.245 & 0.149  & 0.084 & 0.004  & 0.003 &  0.056 &  0.225 & 0.113 & 0.030  & 0.022 &  0.000 & 1.000   &  \\  
\hline\hline
\end{tabular}
  \end{center}
\end{table*}

\section{Results}
\label{sec:res}

As described above we consider several variants when evaluating the muon capture rate. Some of them are
preferable, but we comment on the others as well.

First, two variants, small and large, of the single particle level set are considered. The larger one seems to be preferable. However,
note the issue of the unbound neutron states discussed in Section \ref{sec:par}. As seen in the Table \ref{tab:FGTT} the
calculated capture rates using the small single particle space are typically $\sim$ 20\% smaller than those in the large
single particle space.
Second, the bound muon wave function can be included  in the factorized form, as the $Z_{eff}^4$ factor, or without factorization.
Again, we consider the variant without factorization preferable. The corresponding capture rates are $\sim$10\% smaller than those
evaluated with factorization.
We also verified that the two prescriptions, described in Subsection \ref{sub:one} and \ref{sub:two} lead to the same results.
This is an important test of our procedures and codes.
Finally, there are two ways of reducing the weak Hamiltonian to its nonrelativistic form. As follows from Table \ref{tab.coef2}
and the results in this Section 
using the present, and preferable, prescription results in capture rates that are (20-30)\% smaller than those based on the traditional
Fujii-Primakoff prescription.

Lets discuss first briefly the energy and multipolarity distributions in the final (Z-1,N+1)
odd-odd final nucleus. Note that some, actually most,
states in the final nucleus are unbound and lead eventually to the emission of one or more neutrons.

\begin{center}
  \begin{figure}[htb]
  \includegraphics[width=0.52\textwidth]{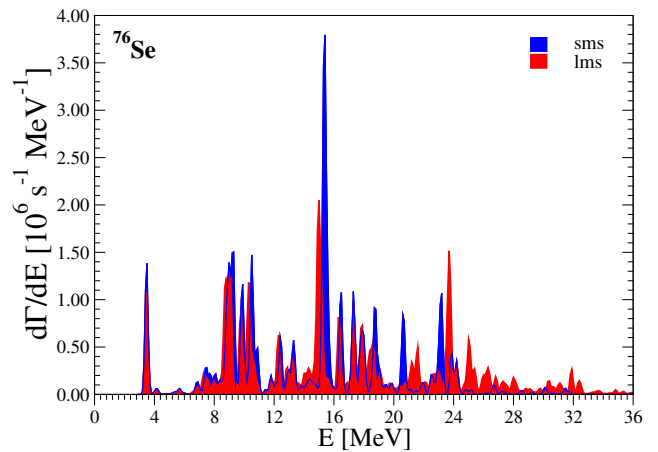}
  \caption{(Color online) Energy spectra of muon capture rate on $^{76}$Se. 
   Results  for  the small (sms, 22 lev.) and the large (lms, 36 lev.)
    single nucleon model spaces are presented.   
}
\label{fig:se-Edis}
\end{figure}
\end{center}

\begin{center}
  \begin{figure}[htb]
  \includegraphics[width=0.52\textwidth]{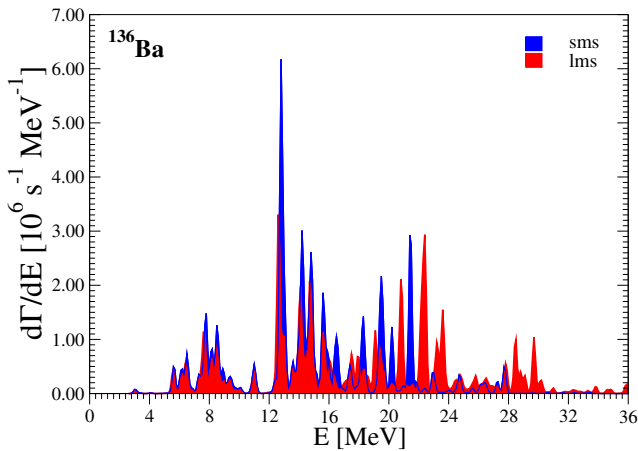}
  \caption{(Color online) Energy spectra of muon capture rate on $^{136}$Ba.
   Results  for  the small (sms, 22 lev.) and the large (lms, 36 lev.)
    single nucleon model spaces are presented.   
}
\label{fig:ba-Edis}
\end{figure}
\end{center}

The energy distributions of the final states in muon capture on $^{76}$Se and $^{136}$Ba
are shown in Figs. \ref{fig:se-Edis} and \ref{fig:ba-Edis}. The results with
small and larger single particle spaces are shown. The discrete final states are replaced
with the Gaussian peaks of 100 keV width. With the larger single particle space not only additional higher
excitation energy states are populated, but the distribution among the lower energy states are also 
noticeably changed. The fraction of bound states below the neutron emission thresholds of  7.33 MeV 
in $^{76}$As is 0.32 for the
large single particle space and 0.36 for the small one. In $^{136}$Cs the neutron emission threshold is 6.83 MeV, and the corresponding
bound state fractions are 0.17 for the large space and 0.21 for the small one.

\begin{center}
  \begin{figure}[htb]
  \includegraphics[width=0.50\textwidth]{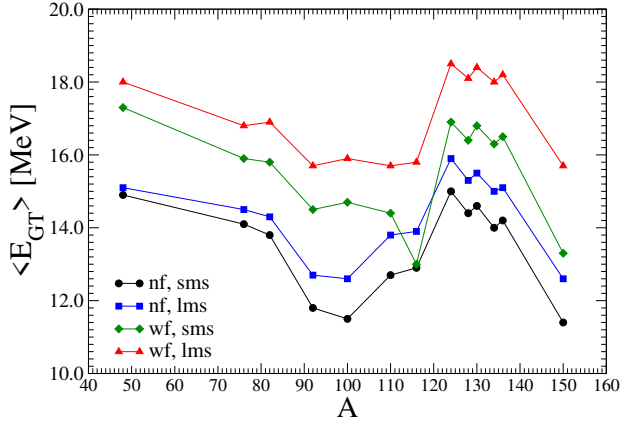}
  \caption{(Color online) The average energys of excited state associated with the Gamow-Teller
    matrix elements $B_{\rm GT}$ and ${\widetilde B}_{\Phi {\rm GT}}$. nf (wf) denotes
    case without (with) factorization of muon wave function and nuclear matrix element.
    sms (lms) stands for small (large) model space calculation.}
\label{fig:averen}
\end{figure}
\end{center}

In Fig. \ref{fig:averen} the average excitation energies associated with the largest GT matrix elements are shown for
all considered nuclei. The shift between the small and large single particle spaces, as well as the considerable shift
between the evaluation with and without the factorization of the bound muon wave function are clearly visible. However,
the pattern of the average excitation energy as a function of the mass number $A$ is very similar in all four variants.

\begin{center}
  \begin{figure}[htb]
  \includegraphics[width=0.52\textwidth]{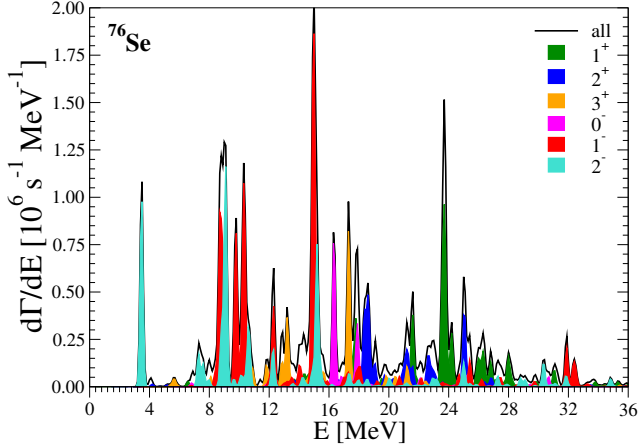}
  \caption{(Color online) Multipole contributions to the energy spectra  of muon capture rate on $^{76}$Se 
     The same notation as in Fig. \ref{fig:se-Edis}
    is assumed. The large single particle  space is considered.}
\label{fig:se-JEdis}
\end{figure}
\end{center}

\begin{center}
  \begin{figure}[htb]
  \includegraphics[width=0.52\textwidth]{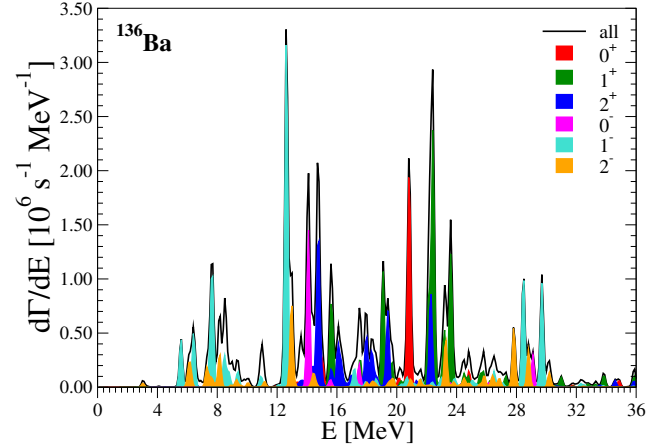}
  \caption{(Color online) Multipole contributions to the energy spectra  of muon capture rate of $^{136}$Ba. 
    The same notation as in Fig. \ref{fig:ba-Edis}
    is assumed. The large single particle  space is considered.
}
\label{fig:ba-JEdis}
\end{figure}
\end{center}

In Figs. \ref{fig:se-JEdis} and \ref{fig:ba-JEdis} the same energy spectra as in Figs. \ref{fig:se-Edis} and \ref{fig:ba-Edis}
are shown, but separated into different multipoles. Only the large single particle model space is used, hence the
scale difference. As is also seen in Table \ref{tab:mult2} the $1^-$, $2^-$, $1^+$ and $2^+$ multipoles dominate, each 
accounting for roughly comparable contributions.

\begin{table}[htb]
  \begin{center}
    \caption{\label{tab:jouni} Comparison of the experimental muon capture rates $\Gamma_{GP}$, based on the empirical
    Goulard-Primakoff \cite{Suz87} formula  with the rate $\Gamma_{QRPA}$ calculated in Ref. \cite{Jok19} and those
    evaluated
    in  this work when using the present $\Gamma_{pres}$ and the Fujii-Primakoff $\Gamma_{FP}$ parametrizations.
    Both  $\Gamma_{pres}$ and $\Gamma_{FP}$ were evaluated using the large single particle model space.
    All calculations use the same $g_A^{\rm eff}$ = 0.8, and the rates are in units of $10^6$/s.  }
\begin{tabular}{lcccc}
  \hline\hline
  nucleus & $\Gamma_{GP}$ & $\Gamma_{QRPA}$ & $\Gamma_{pres}$ & $\Gamma_{FP}$  \\
  \hline
  $^{76}$Se  & 7.00  & 16.4  &  3.50  &  4.66  \\
  $^{82}$Kr  & 7.22  & 16.5  &  3.76  &  5.00  \\
  $^{96}$Mo  & 9.90  & 20.4  &  5.32  &  7.06  \\
  $^{100}$Ru & 11.2  & 16.7  &  5.77  &  7.65  \\
  $^{116}$Sn & 12.7  & 15.7  &  6.61  &  8.73  \\
  $^{128}$Xe & 12.4  & 21.2  &  6.37  &  8.39  \\
  $^{130}$Xe & 11.1  & 23.6  &  5.97  &  7.87  \\
  $^{136}$Ba & 11.1  & 21.1  &  7.61  &  10.0   \\
  \hline
   \hline\hline
  \end{tabular}
  \end{center}
\end{table}

As mentioned in the Introduction, there seems to be a discrepancy between the calculated muon capture rates based on the
QRPA method in the older Refs. \cite{Zin06, Mar09} where none, or only mild, quenching of $g_A$ was required
and more recent Ref. \cite{Jok19} where rather substantial quenching is indicated.
To address this discrepancy explicitly using the present method of calculation, we compare in Table \ref{tab:jouni} the calculated
rates in Ref. \cite{Jok19} and here, with both ways of choosing the constants in the nonrelativistic Hamiltonian.
The experimental capture rates $\Gamma_{GP}$, empirically adjusted for the individual isotopes, are also shown.
Our results in columns 4 and 5 use, for this purpose only, the same $g_A^{\rm eff}$  = 0.8 as in Table II of \cite{Jok19}.  
 One can see that the muon capture rates in Ref. \cite{Jok19} are 2-3 times faster that in our work. 
 At the same time, obviously, the $\Gamma_{QRPA}$ of Ref. \cite{Jok19} are substantially larger than experiment, thus requiring
 even smaller $g_A^{\rm eff}$. Our results are smaller than the  experiment, thus requiring $g_A^{\rm eff} >$ 0.8. The origin of the
 discrepancy is unknown at the present time. 
 
\begin{center}
  \begin{figure}[htb]
  \includegraphics[width=0.50\textwidth]{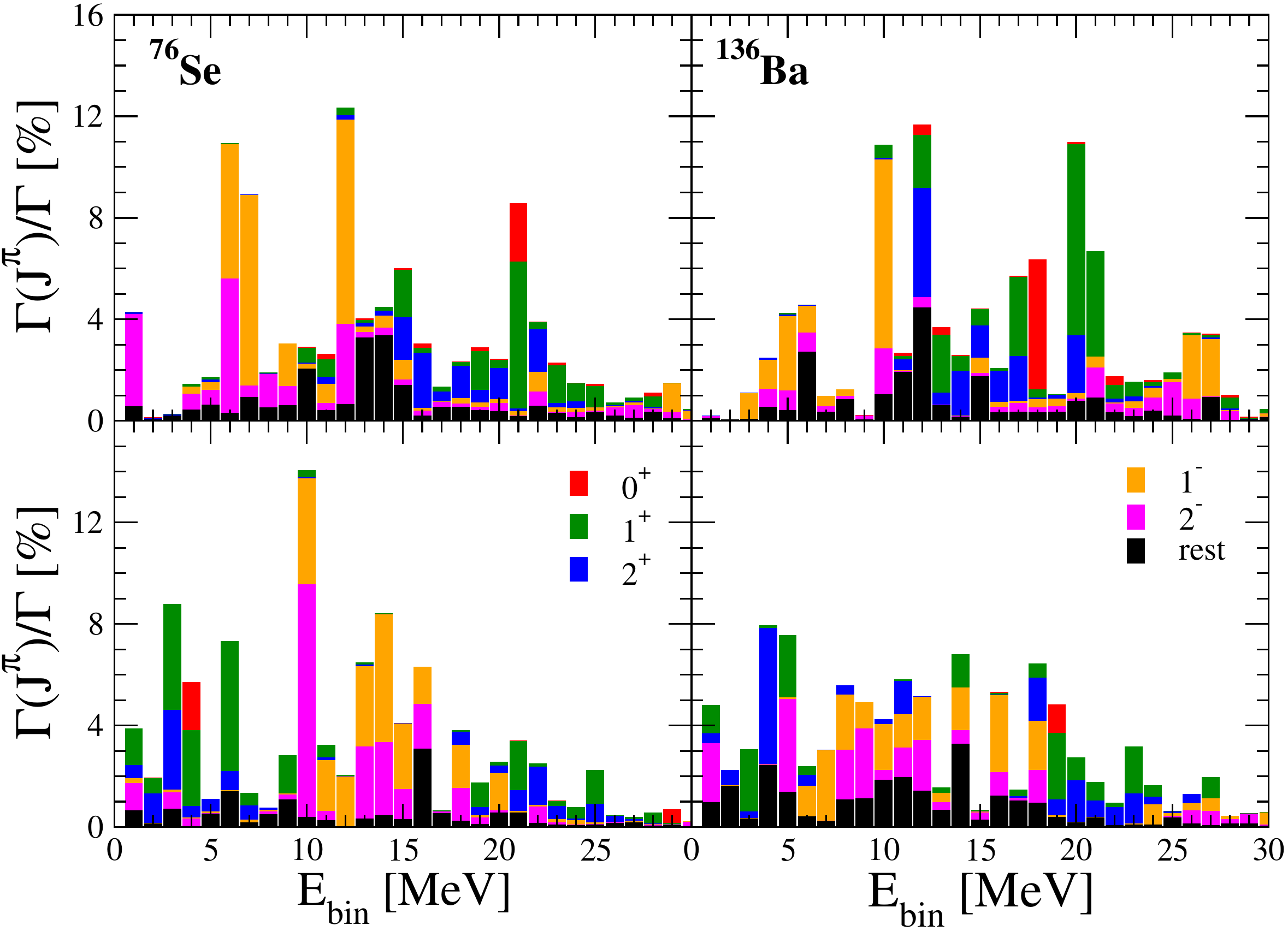}
  \caption{(Color online) Multipole and energy distributions  of the muon capture rate of 
  $^{76}$Se and $^{136}$Ba expressed as fractions of the total capture rate. In the upper
  panels are our results, evaluated using the large single particle model space. In the lower
  panels are the results of Ref. \cite{Jok19}. The energy scale refers to the excitation energy
  in the final odd-odd nuclei.
}
\label{fig:seba-JEdis}
\end{figure}
\end{center}

To see the differences of the results here and in Ref. \cite{Jok19} more clearly 
we compare in Fig. \ref{fig:seba-JEdis} the results of both works 
for $^{76}$Se and $^{136}$Ba used as examples. The differences in both
 the energy and multipole distributions are quite noticeable. It appears
that the present approach leads to somewhat less strength at lower excitation energy and correspondingly more
strength at higher energies compared to the results of Ref. \cite{Jok19}.

Note, that the experimental data are mostly for elements, not for individual isotopes. Thus, instead of using them
directly, we use for comparison with calculations 
the so called Goulard-Primakoff empirical formula \cite{Suz87} that describes sufficiently well the muon capture rate for all nuclei with
given $A$ and $Z$,
\begin{eqnarray}
  &&\Gamma^{\mu A}_{\rm GP}(A,Z) = Z^4_{\rm eff} G_1 \left( 1 + G_2\frac{A}{2 Z} \right.\nonumber\\
  &&~~~~~~~\left. - G_3 \frac{A-2Z}{2 Z} -
G_4 \left( \frac{A-Z}{2 A} + \frac{A- 2 Z}{8 A Z}\right)\right)\nonumber\\  
\end{eqnarray}
where $G_1=261$, $G_2=-0.040$, $G_3=-0.26$, $G_4=3.24$ (TRIUMF data fit). While the agreement of the 
Goulard-Primakoff empirical formula with the few measured capture rates for individual isotopes is not perfect,
the discrepancies never exceed $\sim$ 10\%. 

\begin{center}
\begin{figure}[htb]   
  \includegraphics[width=0.50\textwidth]{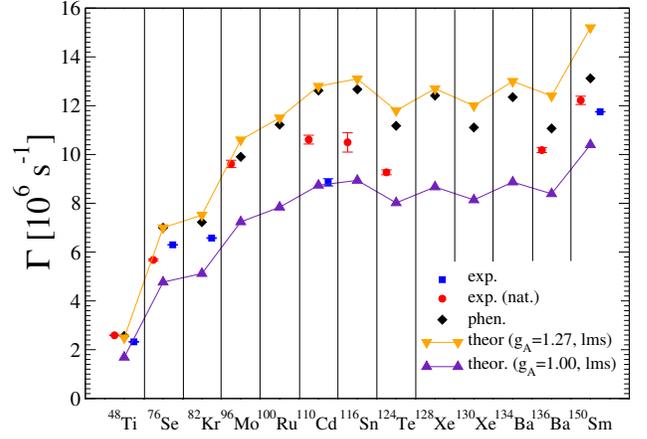}
  \caption{(Color online) A comparison of experimental and theoretical
    total capture rates for the final nuclei participating in the
    double-beta decay. Measurements were performed for 
    a for a given isotope in \cite{Zin19} and for elements 
   with the natural abundance of isotopes in \cite{Mea01}. Theoretical
    results were obtained with large model space for $g_A = 1.00$ and $1.27$ 
}
\label{fig:expth} 
\end{figure}
\end{center}

Our main results  are shown in Fig. \ref{fig:expth} where the experimental total capture rates are compared with the calculated
rates for $g_A^{\rm eff}$ = 1.27 and 1.0, and for all final nuclei participating in the double beta decay. Clearly, for these results, obtained
with the large single particle model space, the experimental data are bracketed by these two $g_A^{\rm eff}$ values. 

More details are shown in Table \ref{tab:exp}. Here the results with and without muon wave function factorization are shown
for both single particle spaces and for $g_A^{\rm eff}$ = 0.8, 1.0 and 1.27. The experimental data for the natural elements from Ref.
\cite{Mea01}  and for separated isotopes from Ref. \cite{Zin19} are also shown for comparison.

\begin{table}[htb]
  \begin{center}
    \caption{\label{tab:exp} The calculated muon capture rates for the final 
      nuclei participating in double beta decay. The rates are evaluated  for $g_A^{\rm eff}$ values 0.80, 1.00 and 1.27,
      as indicated.
      They are obtained with the present
      approach with (with fact.) and without (no fact.) factorization of muon wave
      functions. The small (s) and large (l)
      single particle level spaces are considered. $\Gamma_{\rm exp.}$ is the experimental 
      total capture rate for the stable elements with  natural abundance of isotopes \cite{Mea01}
      and for a given isotope \cite{Zin19}.
      All values of capture rates are in units $10^6~s^{-1}$.}
\begin{tabular}{lcccccccccc}
\hline\hline
nuclide  & $\Gamma_{\rm exp.}$ & & isotope & $g_A^{\rm eff}$ &
\multicolumn{5}{c}{ $\Gamma_{\rm pres}$} & \\ \cline{6-10}
   &   & &                  &       & \multicolumn{2}{c}{ with fact.} & & \multicolumn{2}{c}{ no fact.} & \\
\cline{6-7}\cline{9-10}
   &                    & &                  &       & s & l & & s & l & \\ \hline
$^{\rm nat}$Ti & $2.590\pm 0.012$   & & ${^{48}{\rm Ti}}$ & 0.80   & 1.08 & 1.32 &  & 0.99 & 1.23 &   \\ 
             & $2.60\pm 0.04$     & &                  & 1.00   & 1.47 & 1.81 &  & 1.35 & 1.69 &   \\ 
$^{48}$Ti     & $2.323\pm 0.015$   & &                  & 1.27   & 2.15 & 2.67 &  & 1.97 & 2.49 &   \\ 
$^{\rm nat}$Se & $5.681\pm 0.037$   & & ${^{76}{\rm Se}}$ & 0.80   & 3.30 & 3.83 &  & 2.94 & 3.50 &   \\ 
             & $5.70\pm 0.05$     & &                  & 1.00   & 4.47 & 5.22 &  & 3.98 & 4.77 &   \\ 
$^{76}$Se     & $6.300\pm 0.004$   & &                  & 1.27   & 6.53 & 7.66 &  & 5.81 & 7.00 &   \\ 
$^{82}$Kr     & $6.576\pm 0.017$   & & ${^{82}{\rm Kr}}$ & 0.80   & 3.40 & 4.10 &  & 3.01 & 3.76 &   \\ 
             &                    & &                  & 1.00   & 4.61 & 5.60 &  & 4.08 & 5.12 &   \\ 
             &                    & &                  & 1.27   & 6.76 & 8.22 &  & 5.96 & 7.52 &   \\ 
$^{\rm nat}$Mo & $9.23\pm 0.07$     & & ${^{96}{\rm Mo}}$ & 0.80   & 4.73 & 5.87 &  & 4.24 & 5.32 &   \\ 
             & $9.614\pm 0.15$    & &                  & 1.00   & 6.39 & 8.01 &  & 5.72 & 7.24 &    \\ 
             & $9.09\pm 0.18$     & &                  & 1.27   & 9.30 & 11.8 &  & 8.32 & 10.6 &    \\ 
Ru           &                    & & ${^{100}{\rm Ru}}$ & 0.80  & 4.92 & 6.43 &  & 4.38 & 5.77 &    \\ 
             &                    & &                   & 1.00  & 6.60 & 8.74 &  & 5.88 & 7.84 &    \\ 
             &                    & &                   & 1.27  & 9.54 & 12.8 &  & 8.50 & 11.5 &    \\ 
$^{\rm nat}$Cd & $10.63\pm 0.11$    & & ${^{110}{\rm Cd}}$ & 0.80  & 6.31 & 7.20 &  & 5.58 & 6.44 &    \\ 
             & $10.61\pm 0.18$    & &                   & 1.00  & 8.59 & 9.79 &  & 7.59 & 8.74 &    \\ 
$^{116}$Cd    & $8.86\pm 0.15$     & &                   & 1.27  & 12.6 & 14.3 &  & 11.1 & 12.8 &    \\ 
$^{\rm nat}$Sn & $10.70\pm 0.14$    & & ${^{116}{\rm Sn}}$ & 0.80  & 6.90 & 7.43 &  & 5.96 & 6.61 &    \\ 
             & $10.44\pm 0.18$    & &                   & 1.00  & 9.30 & 10.1 &  & 8.03 & 8.94 &    \\ 
             & $10.5\pm 0.4$      & &                   & 1.27  & 13.5 & 14.7 &  & 11.7 & 13.1 &    \\ 
$^{\rm nat}$Te & $ 9.27\pm 0.10$    & & ${^{124}{\rm Te}}$ & 0.80  & 5.76 & 6.47 &  & 5.03 & 5.89 &    \\ 
   & $ 9.06\pm 0.11$    & &                   & 1.00  & 7.83 & 8.83 &  & 6.83 & 8.02 &    \\ 
   &                    & &                   & 1.27  & 11.5 & 13.0 &  & 9.99 & 11.8 &    \\ 
Xe &                    & & ${^{128}{\rm Xe}}$ & 0.80  & 6.05 & 7.03 &  & 5.26 & 6.37 &    \\ 
   &                    & &                   & 1.00  & 8.22 & 9.60 &  & 7.14 & 8.67 &    \\ 
   &                    & &                   & 1.27  & 12.0 & 14.1 &  & 10.5 & 12.7 &    \\ 
   &                    & & ${^{130}{\rm Xe}}$ & 0.80  & 5.45 & 6.53 &  & 4.86 & 5.97 &    \\ 
   &                    & &                   & 1.00  & 7.56 & 8.93 &  & 6.61 & 8.14 &    \\ 
   &                    & &                   & 1.27  & 11.1 & 13.2 &  & 9.70 & 12.0 &    \\ 
$^{\rm nat}$Ba & $10.18\pm 0.10$    & & ${^{134}{\rm Ba}}$ & 0.80  & 5.88 & 7.19 &  & 5.09 & 6.50 &    \\ 
   & $ 9.94\pm 0.16$    & &                   & 1.00  & 8.03 & 9.84 &  & 6.93 & 8.87 &    \\ 
   &                    & &                   & 1.27  & 11.8 & 14.5 &  & 10.2 & 13.0 &    \\ 
   &                    & & ${^{136}{\rm Ba}}$ & 0.80  & 5.43 & 8.11 &  & 4.72 & 6.14 &    \\ 
   &                    & &                   & 1.00  & 7.43 & 10.6 &  & 6.45 & 8.39 &    \\ 
   &                    & &                   & 1.27  & 11.0 & 14.6 &  & 9.48 & 12.4 &    \\ 
$^{\rm nat}$Sm & $12.22\pm 0.17$    & & ${^{150}{\rm Sm}}$ & 0.80  & 6.34 & 8.23 &  & 5.69 & 7.61 &    \\ 
$^{150}$Sm    & $11.75\pm 0.07$    & &                   & 1.00  & 8.62 & 11.2 &  & 7.72 & 10.4 &    \\ 
   &                    & &                   & 1.27  & 12.6 & 16.5 &  & 11.3 & 15.2 &    \\
\hline\hline
\end{tabular}
  \end{center}
\end{table}

\begin{center}
  \begin{figure}[htb]
  \includegraphics[width=0.50\textwidth]{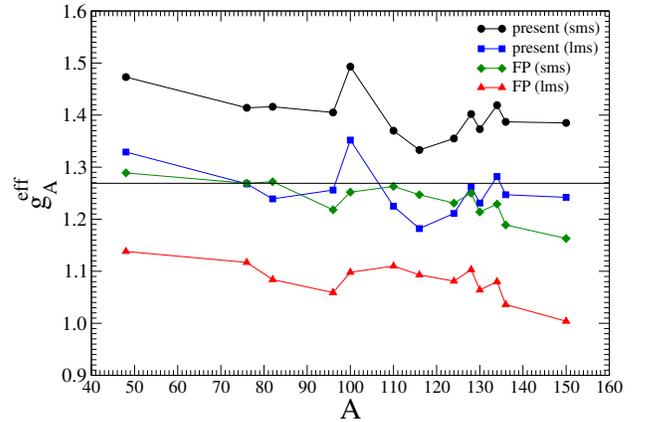}
  \caption{(Color online)
    Effective axial-vector coupling constant $g_A^{\rm eff}$ needed to reproduce the empirical 
    Goulard and Primakoff muon capture rate $\Gamma_{\rm GP}$. Calculations were performed for 
    the same nuclei as in Fig. \ref{fig:expth}.  Results shown are for the present
    and Fujii-Primakoff approaches of reducing the weak Hamiltonian to its nonrelativistic form.
     Both  the small (sms) and large (lms) single particle
    spaces were used. 
}
\label{fig:gaeff}
\end{figure}
\end{center}

Finally, in Fig. \ref{fig:gaeff} we display the values of the effective axial current coupling constant $g_A^{\rm eff}$ needed to 
obtain the empirical total muon capture rate as given by the Goulard-Primakoff formula. It is encouraging to note that
for the preferred variant with the large single particle space and present prescription of reducing the weak Hamiltonian
to its nonrelativistic form (blue line) no quenching at all is required. But given the approximations involved this level
of agreement is perhaps somewhat accidental.  However, the figure shows clearly that no matter what $g_A^{\rm eff} \ge 1.0$
is required to reproduce the experimental muon capture rates.

\section{Conclusions}

The study of muon capture on nuclei is an important test of the ability of nuclear models to describe this semileptonic weak
process. Muon capture is characterized by the relatively large momentum transfer of the order of muon mass, and hence involves many
multipolarities and a wide range of excitation energies. 
The Quasiparticle Random Phase Approximation is a method designed to describe collective nuclear effects.
Thus, as a test of the method, the total muon capture rate is, we believe, 
a characteristic that should be considered first, in preference to the
description of the individual final states that each represent only a small fraction of the total capture rate. 

In this work we show that the QRPA method is capable to describe 
 the muon capture rate in agreement with experiment in a many nuclei, spanning
a large interval of $Z$ and $A$. To reach such a conclusion we used a variety of procedures. Some of them have been
used before, some others are new. It is important to note that the resulting calculated capture rates are 
relatively close to each other, independently of the variant employed. 
That shows that they are relatively stable. It is also an important test of our procedures, since many of the variants require
separate, and seemingly quite different, computer codes.

In particular, our aim is to test whether the idea of the axial current quenching is needed to describe the muon capture.
If it would be the case, we would expect that the calculated capture rates would exceed the experimental ones by an
approximately constant factor for a large group of nuclei. Our results show that, at least for the QRPA method as described
here, this is not the case. We describe the capture rates reasonably well with the standard value of $g_A = 1.27$.   
There is no necessity of any quenching.

More details, like the fraction leading to the bound states in the $(Z-1,N+1)$ nucleus, or the branching ratios for the individual
bound states, might be also eventually used as additional tests of the model.

\begin{acknowledgments}
This work was supported by the VEGA Grant Agency of the Slovak Republic under Contract No. 1/0607/20 and 
by the Ministry of Education, Youth and Sports of the Czech Republic under the  INAFYM Grant 
No.~CZ.02.1.01/0.0/0.0/16\_019/0000766.
The work of P.V. is supported by the Physics Department, California Institute of Technology.
\end{acknowledgments}

\end{document}